\documentclass[bibyear]{aa}

\usepackage{txfonts}
\usepackage{graphicx}
\usepackage{natbib}
\bibpunct{(}{)}{;}{a}{}{,} 
\newcommand{\htwoo}{H$_2$O}
\newcommand{\methanol}{CH$_3$OH}
\newcommand{\Lsun}{L$_\odot$}

\newcommand{\s}{s$^{-1}$}
\begin{document}

   \title{The OH Masers Towards IRAS~19092+0841}


   \author{K.A. Edris
          \inst{1,2}
          \and
          G.A. Fuller\inst{1}
          \and
          S. Etoka\inst{3}
          \and
          R.J. Cohen\inst{1}
          }

\institute{Jodrell Bank Centre for Astrophysics, School of Physics and Astronomy, Alan Turing Building University of Manchester, Manchester, M13 9PL, UK
\and
 Al-Azhar University, Faculty of Science, Astronomy Department, PO Box 11884, Naser City, Cairo, Egypt\\
 \email{khedres@azhar.edu.eg}
\and
 Hamburger Sternwarte, Gojenbergsweg 112, 21029 Hamburg, Germany
 }

   \date{Received ; accepted by Astronomy and Astrophysics }


  \abstract
   {Maser emission is a strong tool for studying high mass star forming regions and their evolutionary stages. OH masers in particular can trace the circumstellar material around protostars and determine their magnetic field strengths at milliarcsecond resolution. }
   {Imaging OH maser mission towards high mass protostellar objects to determine their evolutionary stages and to locate the detected maser emission in the process of high mass star formation. }
   {In 2007, we surveyed OH maser towards 217 high mass protostellar objects to study its presence. In this paper, we present a follow up MERLIN observations of a ground state OH maser emission towards one of these objects, IRAS 19092+0841. }
   {Emission from the two OH main spectral lines, 1665 and 1667 MHz, were detected close to the central object. The positions and velocities of the OH maser features have been determined. The masers are distributed over a region of $\sim 5''$ corresponding to 22400 AU (or $\sim$ 0.1 pc) at a distance of 4.48 kpc. The polarization properties of the OH maser features were determined as well. We identify three Zeeman pairs from which we inferred a magnetic field strength of $\sim 4.4 mG$ pointing towards the observer.}
   { The relatively small velocity spread and the relatively wide spacial distribution of the OH maser features support the suggestion that this object could be in an early evolutionary state before the presence of disk and/or jet/outfows.}

\titlerunning{The OH Masers Towards IRAS~19092+0841.}
\authorrunning{Edris et al.}

   \maketitle
%

\section{Introduction}

   Observations show that OH masers are associated with different evolutionary stages of star forming regions. It was believed that they are associated with only HII regions (e.g. Garay \& Lizano \citealp{garay99}). However observations which have been carried out towards a number of star forming regions show that OH masers are also associated with an earlier stage before the appearance of ionized HII regions (e.g. Cohen et al. \citealp{cohen88}; Braz et al. \citealp{braz90}). This later type of OH maser is associated with an accretion phase, outflow, and circumstellar disks (Brebner et al. \citealp{brebner87}; Hutawarakorn, \& Cohen \citealp{hutawarakorn99}; Hutawarakorn et al. \citealp{hutawarakorn02}; Fuller et al. \citealp{fuller01}; Edris et al. \citealp{edris05}). An OH maser survey by Edris, Fuller, \& Cohen (\citealp{edris07}, hereafter EFC07) detected the OH masers towards 26\% towards a sample of 217 High Mass Protostellar Objects (HMPOs) candidates. It is of interest to know the distributions of the OH masers and their associations towards these regions in addition to their position in the evolutionary sequence. Maser emission gives a unique opportunity of observing those typically far regions in some details. This paper is the second follow up of EFC07 survey with a high angular resolution observation. One of EFC07 objects (IRAS 20126+4104, hereafter IRAS20126) was studied by Edris et al. (\citealp{edris05}) in the ground-state OH, 22-GHz \htwoo, and 6.7-GHz class II \methanol\ masers transitions/lines.

The present object of study, IRAS~19092+0841 (hereafter IRAS19092), is one of the HMPOs sample studied by Palla et al. \cite{palla91} and Molinari et al. (\citealp{molinari96}; \citealp{molinari98}; \citealp{molinari00}). Palla et al. \cite{palla91} divided this sample of 260 IRAS sources into two sub-samples of '\textit{high}' and '\textit{Low}' sources. The \textit{High} sources are the sources located in the 'higher' part of the color-color diagram such that: [25-12]$>=$0.57, which in essence is the prescription for the presence of associated UCHII regions according to Wood \& Churchwell \cite{wood89} criteria. On the other hand, the '\textit{Low}' sources are most probably made of two groups of sources with a distinct evolutionary state (Molinari et al. \citealp{molinari96}). The first '\textit{Low}' group of sources is believed to be at a very '\textit{early}' evolutionary state before the creation of a 'real' YSO at the centre and consequently before the appearance of any measurable/detectable UCHII regions. The other '\textit{Low}' group of sources is believed to contain more evolved sources (older than the '\textit{High}' sources) containing objects which have already dispersed much of their circumstellar material. IRAS19092 belongs to the latter group of the '\textit{Low}' sub-sample. IRAS19092 is at an estimated distance of 4.48 kpc (Molinari et al. \citealp{molinari96}) and has a luminosity of $10^4$ \Lsun. An ammonia core was found towards IRAS19092 at velocity $\sim$ 58 km~\s\ (Molinari et al. \citealp{molinari96}).

IRAS19092 is associated with different types of maser emission. OH maser was originally reported by MacLeod et al. \cite{macleod98} and then by EFC07. Water maser was detected by Palla et al. (\citealp{palla91}). IRAS19092 is also associated with the class I methanol masers emission at 44-GHz, but it is not associated with maser emission at 95-GHz (Kurtz, Hofner, \& Alvarez \citealp{kurtz04}; Fontani et al. \citealp{fontani10}). Class II \methanol\ masers at 6.7 GHz were detected by Szymczak et al. (\citealp{szymczak00}) and mapped by Pandian et al. \cite{pandian11}. CO and H$_2$ observations failed to detect any sign of outflows in this region (Zhang et al. \citealp{zhang05}; Varricatt et al. \citealp{varricatt10}). It is not associated with a close cm radio continuum emission. The closest cm continuum emission detected by Molinari et al. (\citealp{molinari98}) at 6 cm is offset from the IRAS position by 110 arcsec. However, It is associated with mm continuum emission detected by Molinari et al. \cite{molinari00} and the Bolocam Galactic Plane Survey (BGPS, Rosolowsky et al. \citealp{rosolowsky10}).

To determine what the OH masers trace and how they are distributed and related to other tracers, IRAS19092 has been observed at high angular resolution using MERLIN. The details of the observations and reduction are given in Sec. 2 and the results presented in Sec. 3. In
Sec. 4 we discuss the interpretation while conclusions are drawn in Sec. 5.
\begin{table}
 \centering
 \caption{Observing and calibration parameters for the MERLIN
spectral-line observations of IRAS19092}
\begin{tabular}{c c}  \hline

  {Observational parameters}  & {OH masers}  \\ \hline

  Date of observation & 4  \& 5 April 2003     \\
  Antenna Used & seven antenna  \\
  Field centre (2000) & $\alpha =  19^{h} 11^{m} 37.40^{s}$ \\
                      & $\delta  = 08^{\circ} 46' 30.00''$  \\
  Rest frequency (MHz) & 1665.402 \\
                       & 1667.359  \\
  No. of frequency channels & 512  \\
  Total bandwidth (MHz) & 0.5 \\
  bandpass calibrator & 3C84  \\
  Polarization angle calibrator & 3C286  \\
  Phase calibrator & 1919+086 \\ \hline
\end{tabular}

\end{table}

\section{Observations and data reduction}

Table 1 gives the parameters for the MERLIN observations. The phase calibrator source 1919+086 was used to retrieve the absolute position of the maser components and therefore compare their locations from one line to another. A bandpass calibrator 3C84 was observed to calibrate the variation of instrumental gain and phase across the spectral bandpass. Observations of 3C286 were also made during the observing run, with the same correlator configuration and bandwidth, to calibrate for the polarization angle.

IRAS19092 was observed in the 1665- and 1667-MHz OH maser transitions in April $2003$ using the seven telescopes of the MERLIN network available then. During the observations, the frequency was cycled between the two OH line frequencies, to provide data on both transitions spread over the whole observing track. The velocity resolution was 0.21 km~\s\ for a total of 0.25 MHz spectrum bandwidth corresponding to a 45 km~\s\ velocity range. The observations were performed in full polarization mode. Edition of obvious bad data and correction for gain-elevation effects were performed using the MERLIN d-programs (see Diamond et al. \citealp{diamond03}). The flux density of the amplitude calibrator 3C84, was determined by comparing the visibility amplitudes on the shortest baselines with those of 3C286. Using flux density of 13.625 Jy for 3C286 (Baars et al. \citealp{baars77}), the flux density of 3C84 at the time of the observation was determined to be 21 Jy.

In AIPS more refine edition was performed and the data were calibrated for all remaining instrumental and atmospheric effects. Starting from a point source model, the phase calibrator was mapped, with a total of three rounds of phase self-calibration and the resulting corrections applied to the source data.  The polarization leakage for each antenna was determined using 3C84 and the polarization position angle correction was performed using 3C286. Maps of the LHC and RHC emission were produced, and finally the maps of the Stokes parameters (I, Q, U and V) were produced, using clean algorithms in aips. The rms noise, after CLEANing, was typically 14 mJy/beam and the FWHM of the restoring beam is 0.9 $\times$ 0.4 arcsec at a position angle of $-31^\circ$.

The individual channel maps showed usually one, two or three unresolved components, and each component was usually seen across several spectral channels. The positions of the maser components were determined by fitting two dimensional Gaussian components to the brightest peaks in each channel map. Components were considered as spectral features if they occurred in three or more consecutive channels and were then grouped into spectral features. The positions
and velocities of these maser features were obtained by taking flux weighted means over those channels showing emission from the feature. The uncertainties in relative positions are given in Table 2, while the absolute positional accuracy is estimated as described in Edris et al. \cite{edris05} to be better than 30 mas. This positional uncertainty depends on four factors, the position accuracy of the phase calibrator, the accuracy of the telescope positions, the relative position error depending on the beamsize and signal-to-noise ratio and finally the atmospheric variability.


\begin{table*}
  \centering
  \caption{The parameters of the left and right hand circular polarization features of 1665-MHz and the 1667-MHz OH masers detected towards
 IRAS19092.}
\begin{tabular}{c c c c c c c c c} \hline

{No.} & {Vel.} &   {Peak intensity} &  {RA}    &  Error  & {DEC} & Error  & Zeeman & B  \\
   &    km \s   &   Jy beam$^{-1}$ &  h ~ m ~ s &  s    &$^{\circ}$~ $'$ ~  $''$      &  $''$  & pair & mG \\ \hline
\\
1665-MHz & & & & &  &  &  & \\
LHC  & & & & & &   &  & \\
1  & 60.65  &  0.17 $\pm$    0.01  &   19    11   39.002  &  0.004   &  8    46     30.38   &   0.12 & Z1 & \textbf{-4.6} \\
2  & 60.59  &  0.77 $\pm$    0.01  &   19    11   38.973  &  0.000   &  8    46     31.10   &   0.01 & Z2 & \textbf{-4.3} \\
3  & 60.70  &  0.05 $\pm$    0.01  &   19    11   38.899  &  0.008   &  8    46     28.33   &   0.18 &  &  \\
4  & 60.58  &  0.18 $\pm$    0.01  &   19    11   38.943  &  0.001   &  8    46     27.68   &   0.03 & Z3 & \textbf{-4.2} \\
5  & 60.50  &  0.10 $\pm$    0.01  &   19    11   38.909  &  0.003   &  8    46     44.80   &   0.06 &\\
RHC  &  & & & & & &  &  \\
6  & 57.98  &  0.48 $\pm$    0.02  &   19    11   39.001  &  0.003   &  8    46     30.44   &   0.07 & Z1 & \\
7  & 58.08  &  1.53 $\pm$    0.02  &   19    11   38.974  &  0.000   &  8    46     31.09   &   0.00 & Z2 & \\
8  & 58.07  &  0.14 $\pm$    0.02  &   19    11   38.941  &  0.006   &  8    46     27.95   &   0.15 &  & \\
9  & 58.09  &  0.25 $\pm$    0.02  &   19    11   38.950  &  0.001   &  8    46     27.59   &   0.02 & Z3 & \\
\\
1667-MHz &  & & & & & &  &   \\
LHC &  & & & & & & &  \\
1  & 60.67  &  0.48 $\pm$    0.01   &  19    11   38.950  &  0.001   &  8    46     27.81   &   0.03 & & \\
2  & 60.55  &  0.40 $\pm$    0.01   &  19    11   38.944  &  0.002   &  8    46     27.66   &   0.04 & & \\ \hline
\end{tabular}\\
\small - \emph{Note that the values in column 5 and 7 refer to the relative position error while the absolute position uncertainty is $\sim$~30 mas (cf. Sec.~2).}
    \label{tab:oh-comp}
\end{table*}

\begin{table*}
  \centering
 \caption{The Stokes and polarization parameters of the 1665- and 1667-MHz OH masers features detected towards IRAS19092.}
    \label{tab:oh-por}

    \begin{tabular}{c c c c c c c c c c c} \hline

{No.}& {Vel.} &    {$I^a$} &  {$Q^b$}     &   {$U^c$}    &     {$V^d$}  &  {$P^e$} &{$\chi^f$}&{m$_L^g$}&{m$_C^h$} & {m$_T^k$}\\
   &km \s&   Jy beam$^{-1}$ &  Jy beam$^{-1}$    &  Jy beam$^{-1}$    &   Jy beam$^{-1}$  & Jy beam$^{-1}$ &$^{\circ}$&\%
   &  \%    &   \%    \\ \hline
\\
1665-MHz &  & & & & & & & & & \\
1  & 60.65   &  0.05   &   0.00    &   0.02   &  -0.04  &    0.02 &   -     &  33.9  &  -81.4  &  88.2 \\
2  & 60.59   &  0.44   &   0.04    &   0.07   &  -0.32  &    0.08 &   30.1  &  18.1  &  -71.2  &  73.5 \\
3  & 60.70   &  0.04   &   0.00    &   0.02   &   0.00  &    0.02 &   -     &  49.2  &    3.8  &  49.3 \\
4  & 60.58   &  0.11   &   0.00    &   0.00   &  -0.05  &    0.00 &   -     &   0.0  &  -45.3  &  45.3 \\
5  & 60.50   &  0.05   &   0.00    &   0.00   &   0.00  &    0.00 &   -     &   0.0  &    0.0  &   0.0 \\
6  & 57.98   &  0.12   &   0.00    &  -0.08   &   0.09  &    0.08 &   -     &  65.6  &   74.0  &  98.9 \\
7  & 58.08   &  1.23   &  -0.12    &  -0.71   &   0.47  &    0.72 &   -40.2 &  58.3  &   38.5  &  69.8 \\
8  & 58.07   &  0.07   &   0.00    &  -0.05   &   0.05  &    0.05 &   -     &  69.9  &   61.7  &  93.2 \\
9  & 58.09   &  0.22   &  -0.03    &  -0.15   &   0.07  &    0.16 &   39.4  &  70.1  &   31.3  &  76.8 \\
\\
1667-MHz &  & & & & & & & & & \\
1  & 60.67   &  0.33   &   0.02    &   0.02   &   0.02  &    0.03 &   22.5  &   9.1  &    6.1  &  10.9 \\
2  & 60.55   &  0.28   &   0.02    &   0.02   &   0.02  &    0.03 &   22.5  &  10.7  &    7.1  &  12.8 \\
\hline
\end{tabular}\\
\small - \emph{
 $^a$~$I(u,v)= 1/2[RR(u,v)+ LL(u,v)]$,
 $^b$~$Q(u,v)= 1/2[RL(u,v)+ LR(u,v)]$,
 $^c$~$U(u,v)= 1/2[LR(u,v)- RL(u,v)]$,
 $^d$~$V(u,v)= 1/2[RR(u,v)- LL(u,v)]$,
 $^e$~$PPOL$ = (Q$^2$ + U$^2$)$^{(1/2)}$,
 $^f$~$PANG = (0.5 \times  arctan (U/Q))\times (180/\pi) $,
 $^g$~$m_L(\%)= 100 \times P/I $,
 $^h$~$m_C(\%)= 100 \times V/I $,
 $^k$~$m_T(\%)= 100 \times (Q^2 + U^2 + V^2)^{(1/2)}/I$
}
\label{tab:oh-polar}
\end{table*}

\begin{figure*}
\centering
  \includegraphics[angle=0,width=12.5cm]{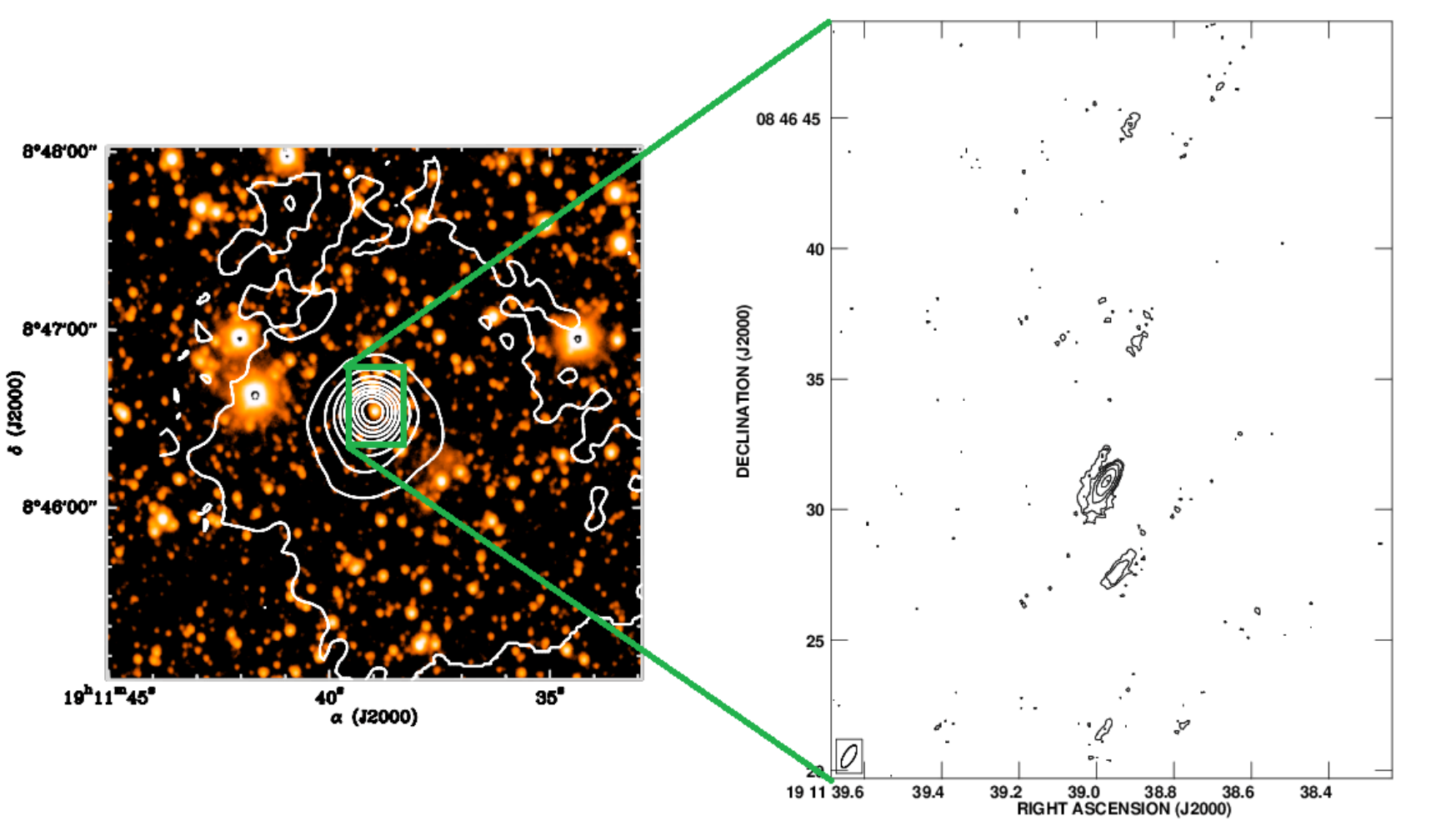}
\caption{\textit{Left panel}: An image for the IRAS19092 region at the Ks band at 2.17 $\mu~m$ with superimposed SCUBA 850 $\mu~m$ continuum in white contours (From Faustini et al. 2009; Fig.A.16). The small box in the middle represent the area of 1665-MHz OH maser maps showed in the \textit{Right panel}. The OH masers peak flux is 7.24 Jy beam$^{-1}$ and the contour levels are 0.12 x (4, 8, 16, 32, 60, 120).
}
    \label{fig:oh_maps}
\end{figure*}

\section{Results}

The 1665- and 1667-MHz OH maser lines were detected with MERLIN. The absolute position of the brightest maser feature in the 1665 MHz OH line is $19^h$ $11^m$ $38.974^s$ $+08^{\circ}$ $46'$ $ 31.09''$ at velocity 58 km~\s. Radial velocities, here and elsewhere, are given relative to the Local Standard of Rest (LSR).

A total of 11 (9 in the 1665-MHz line and 2 in the 1667-MHz line) OH maser features were detected. At the 1665-MHz line, there are 5 left hand circular polarization (LHC) and 4 right hand circular polarization (RHC) features while the two features detected at the 1667-MHz line are LHC. Table \ref{tab:oh-comp} presents the parameters of the OH maser features detected, namely the velocities, peak intensities and positions for each hand of circular polarization. The label Z identifies the left-hand and right-hand polarized features of a Zeeman pair. These Zeeman pairs were identified by searching Table \ref{tab:oh-comp} and \ref{tab:oh-por} for groups of features of opposite polarization that coincide to within the positional uncertainties. Three possible Zeeman pairs were identified which indicates that the 1665 MHz OH line in IRAS19092 suffer zeeman splitting of several km~\s\ exceeding the linewidth (less than 1 km~\s). The strength of the magnetic field can be measured from the velocity difference between the two hands of polarization (Elitzur \citealp{elitzur96}). The splitting of $+2.49$ to $+2.67$ km~\s\ between the left- and right-hand polarized features of the 3 Zeeman pairs Z1-Z3 (cf. Table~2) leads to a magnetic field strengths ranging from 4.2 to 4.6 mG pointing towards us. The magnetic field measured towards IRAS19092 is quite similar to what was measured towards other SFRs such as W3(OH) (Garcia-Barreto et al. \citealp{garcia88}), W75N (Hutawarakorn, Cohen \& Brebner, \citealp{hutawarakorn02}) and W51 (Etoka, Gray \& Fuller, \citealp{etoka12}) harbouring massive central objects at early stages of the star-forming process.

Figure~\ref{fig:oh_maps} shows the map of the OH maser compared to the 1.1 mm map from Faustini et al. \cite{faustini09}. The OH maser are spread over a region of $\sim5''$ (or 17$''$ if the far northern 1665-MHz feature 5 is included) corresponding to 22400 AU (or $\sim$ 0.1 pc) at a distance of 4.48 kpc. The two features of the 1667-MHz line (Table \ref{tab:oh-comp}) are closely associated  ($\le 300$~mas) with the lowest-declination components at 1667~MHz of relatively weak intensity.

Table \ref{tab:oh-por} presents the Stokes parameters I, Q, U and V, the polarization position angle $\chi$ (angles are measured from North towards East), the linearly polarized flux P, the percentage of linear polarization $m_L$, the percentage of circular polarization $m_C$ and the total percentage of polarization $m_T$ of each feature. The Stokes intensities are shown as zero in this table if their flux is below the noise level. There are only two features which are not polarized.

\begin{figure}

  \centering
  \includegraphics[angle=0,width=8cm]{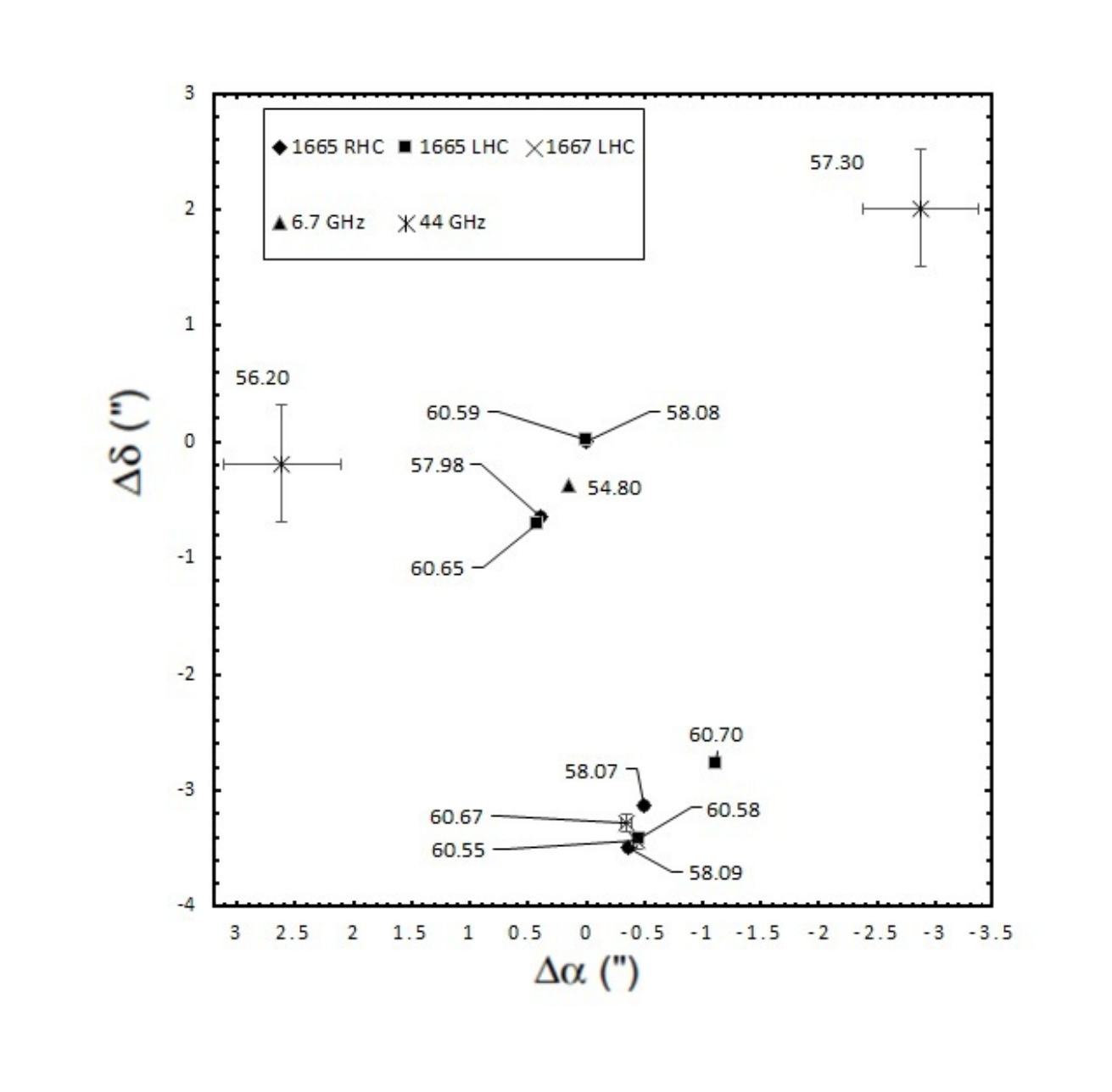}
\caption{ The positions of the MERLIN 1.6-GHz OH maser features, VLA 44-GH class I Methanol maser features from Kurtz et al. (\citealp{kurtz04}) and MERLIN 6.7 GHz class II methanol masers (Pandian et al. \citealp{pandian11}). The symbol which refers to each tracer is shown in the upper left corner of the plot. Note the close association between the OH masers and some 6.7-GH Methanol maser features. The absolute positional accuracies of the MERLIN 1.6-GHz and 6.7-GHz masers are 30 mas (see section 2) and 15 mas respectively, which is smaller than the size of the symbols. The VLA absolute positional accuracies in RA and DEC associated with the 44-GHz Methanol masers are shown by the horizontal and vertical bars respectively. The origin is at RA (J2000)~$=$~$19^h$ $11^m$ $38.974^s$, Dec. (J2000)~$=$~ $+08^{\circ}$ $46'$ $ 31.09''$. The velocities of the maser features are given. For convenient, the fifth feature of the LHC ($\sim~14"$ offset) and a far north feature of the 44-GHz ($\sim~7"$ offset) were not included in this figure.}
\label{fig:pos}
\end{figure}

\section{Discussion}

  The observations reported towards the IRAS19092 region suggest that it is a star forming region in an early evolutionary state before the forming of HII region. It is not associated with neither cm continuum emission nor outflow signature (Molinari et al. \citealp{molinari98}; Zhang et al. \citealp{zhang05}; Varricatt et al. \citealp{varricatt10}). The velocity spread and distribution of the OH maser emission of the observations presented here may indicate that collapsing processes are dominant. The velocity extent is less than 3 km~\s\ for an overall maser emission extent of 22400 AU. Wu et al. \cite{wu07} reported a close $\sim~10''$ south-west source with some spectral lines showing a blue profile expected to be the signature of an inflow motion. The weak flux density of the detected OH masers, suggests that IRAS19092 is in an early evolutionary state. The survey of EFC07 for OH masers towards high mass protostellar objects found that most of the detected sources show OH flux densities $<$ 4.5 Jy, while the OH masers associated with HII regions show much higher flux densities (e.g. Gaume, \& Mutel \citealp{gaume87}; Gasiprong, Cohen, \& Hutawarakorn \citealp{gasiprong02}). Fontani et al. (\citealp{fontani06}) reported the lack of deuterium towards this region and concluded that IRAS 19092+0841 is associated with a cold and dense gas with chemical and physical conditions identical to those associated with low mass starless cores. The presence of class I methanol masers (Kurtz, Hofner, \& Alvarez \citealp{kurtz04}) support the suggestion of early evolutionary state. Ellingsen (\citealp{ellingsen06}) suggested that the class I methanol masers may signpost an earlier stage of high-mass star formation than the class II masers. The early evolutionary stage interpretation is also consistent with the suggestion of Molinari et al. (\citealp{molinari96}) that some of the sources in the \textit{Low} sample (which IRAS19092 is related to, see Sec.~1) are in an earlier evolutionary state than their counterparts in the \textit{High} sample.

\subsection{Comparison with other tracers}
\label{sec:comp}

The OH masers were previously detected by Macleod et al. (\citealp{macleod98}) and EFC07. There is no remarkable change between the three observations in the spectrum, velocity range or peak velocity but the flux density in the later survey of EFC07 was relatively stronger (3.45 Jy). Also in the survey of EFC07 the 1665 MHz (at LHC) spectrum showed a weak feature at velocity 55 km~\s~ which did not appear in the spectrum of Macleod et al. (\citealp{macleod98}) or in the present observations. This feature at 55 km~\s~ has the same velocity as the class II methanol maser at 6.7 GHz detected by Szymczak et al. (\citealp{szymczak00}) and more recently by Fontani et al. (\citealp{fontani10}). In the 6.7 GHz spectrum of Fontani et al. (\citealp{fontani10}) and Pandian et al. \cite{pandian11} a new weaker methanol maser feature appears at velocity $\simeq$ 63 km~\s~(their Figure A-2 and Figure 1).

 The velocities of the OH maser features are more consistent with those of the 44 GHz class I than those of the 6.7 GHz class II methanol masers. On the other hand, the 6.7 GHz class II methanol maser component (centered at V$\sim$~55 km~\s) do not coincide with any OH maser spectral features. The close association of OH (in particular the 1665 MHz line) and class II methanol masers have been proposed by Caswell (\citealp{caswell96}) from subarcsec accuracy survey and modeled by Cragg, Sobolev, \& Godfrey. (\citealp{cragg02}). In some cases a disk has been suggested to be the source of the two maser types (Edris et al. \citealp{edris05}; Gray et al. \citealp{gray03}). However, a close association between OH masers and class I methanol masers towards sources in such early evolutionary state has never been reported so far. The flux ratio of these masers, S(6668)/S(1665)= 6/1.5 = 4 and S(44)/S(1.6) = 1/1.5 = 0.6, places IRAS19092 in OH-favored sources. Note that although the OH and class I \methanol\ masers are associated, there is a clear difference in position, amounting to 2.7 arcsec ($\sim$ 0.06 pc), suggesting that the OH and class I \methanol\ masers are not co-propagating. The water masers associated with IRAS19092 (Palla et al. 1991; Brand et al. \citealp{brand94}) peak at velocity $\simeq$ 57 km~\s~ with a velocity range of 2.5 km~\s. This may indicate a more compact region than that of the OH masers.

The velocity of the strongest OH maser feature agrees with the gas velocity of the ammonia core measured by Molinari et al. (\citealp{molinari96}) as well as the peak velocity of the C$^{34}$S observations (which is assumed to represent the velocity of the high-density gas) carried out by Brand et al. (\citealp{brand01}). This indicates that the OH masers emission originated from the core of this region which is also consistent with the submm map of Faustini et al. (\citealp{faustini09}, Figure 1).

 The OH masers seem to arise from the core of the 850~$\mu$m continuum emission. However the relatively small velocity spread of the OH maser features despite their relatively large distributions of 5$''$ (or 17$''$ if the far northern feature [1665-MHz F5] is included) indicates that the maser emission arises from material which is not close to the central object. The relatively small velocity spread may also refer to weak angular momentum and collapsing. This also indicates that the OH maser emission arises from material which is not very close to the central object. This is consistent with the relatively weak excitation temperature of 10 K (Fontani et al. \citealp{fontani06}) compared to the gas kinetic temperature of 40 K (Brand et al. \citealp{brand01}). Measuring the deuterium fractionation and the CO depletion factor of IRAS19092 among 10 high-mass protostellar candidates, Fontani et al. (\citealp{fontani06}) proposed two scenarios for the location of emitting gas: 1) the cold gas is distributed in an external shell not yet heated up by the high-mass protostellar object, a remnant of the parental massive starless core. 2) the cold gas is located in cold and dense cores close to the high-mass protostar but not associated with it. The observations presented here support the first scenario mentioned. This is also consistent with the association of the OH masers with the class I methanol masers at 44-GHz detected by Kurtz, Hofner, \& Alvarez (\citealp{kurtz04}). If the second scenario is true then the driving source needs still to be identified. Potential candidates are the two faint near-IR sources (2MASS K-Band survey) and the 1.1 mm sources (BGPS Survey, figure \ref{fig:pos}). Unfortunately, the poor positional uncertainty of these tracers does not allow us to draw any conclusion.

It is unclear whether the water and class II methanol masers at 6.7 MHz are associated with another source or associated with the driving source. The peak velocities and velocity ranges of these other maser types are slightly different from those of 1.6-GHz OH masers. The water maser emission peak is centered at V $=$ 57.32~km~\s\ according to Palla et al. \cite{palla91} single-dish observations, while the 6.7 GHz methanol maser emission peak is centered at V $=$ 55~km~\s\ (Pandian et al. \citealp{pandian11}). High angular resolution observations of the water maser emission are needed in order to compare its location with respect to the other maser species in the region and infer what it traces.

\subsection{Comparison with IRAS20126}

IRAS20126 has been mapped at same high angular resolutions by Edris et al. (2005) in the ground-state lines of OH masers as well as the 22-GHz H$_2$O and the 6.7-GHz (class II) CH$_3$OH masers. Comparing the velocity range of the OH masers of IRAS19092 with that of IRAS20126, it is clear that towards the later the velocity range ($\sim$ 17 km~\s) spread much more than that of the former ($\sim$ 3 km~\s) although the spatial distributions of the maser features work oppositely. The OH maser features of IRAS19092 trace an area of angular size three times that of IRAS20126. The OH masers in IRAS20126 trace a circumstellar disk while there is no signature of a circumstellar disk towards IRAS19092. The strength of the magnetic field measured by a Zeeman pair in IRAS20126 is approximately 3 times stronger than that of IRAS19092. This is also consistent with the classification of Molinari et al. (1996). They classify IRAS19092 as a source from the \textit{Low} sample while IRAS20126 belongs to the \textit{High} sample.

The association of OH masers and the two classes of methanol masers are common between these two sources. Towards IRAS20126, the class I 44 GHz methanol masers imaged at arcsec resolution using the VLA by Kurtz et al. (\citealp{kurtz04}) are $\sim$ 7 arcsec ($\simeq$ 0.05 pc) away from the OH maser position while the class II 6.7 GHz methanol masers are just $\simeq$ 0.1 arcsec ($\simeq$ 0.001 pc or 170 AU) away. Towards IRAS19092, the closest feature of the class I 44 GHz methanol masers by Kurtz et al. (\citealp{kurtz04}) is $\simeq$ 0.2 arcsec (0.004 pc) away from the OH masers, while the class II 6.7 GHz methanol masers are $\simeq$ 0.4 arcsec ($\simeq$ 0.009 pc) away.

Towards IRAS20126, the flux density of the OH masers is stronger than that of IRAS19092. Only the 1665-MHz OH line has been detected towards IRAS20126 while the 1665- and 1667-MHz lines have been detected towards IRAS19092. The presence of the two OH mainlines towards IRAS19092 indicates lower gas temperatures and lower density according to the models of Cragg et al. (\citealp{cragg02}) and Gray et al. (\citealp{gray91}). All this indicates that IRAS19092 is in an earlier evolutionary stage than IRAS20126.

\section{Conclusions}

  Ground state OH maser emission at 1665 and 1667~MHz was observed towards IRAS19092 at high angular resolution using MERLIN. Most of the OH maser features are spread over a region of $\sim5''$ corresponding to 22400 AU (or $\sim$ 0.1 pc) at a distance 4.48 kpc. We identify three Zeeman pair, indicating a magnetic field strength ranging from $\sim4.2$ to $\sim4.6$ mG. The absence of any sign of disk, outflow or HII region, and the presence of 44-GHz class I methanol masers suggests that this source is in a very early evolutionary stage of star formation. High angular resolution observations of other tracers in the region are needed  to study this SFR in more depth.

\vspace{1cm}

ACKNOWLEDGMENTS

MERLIN is a national facility operated by the University of Manchester at Jodrell Bank Observatory on behalf of PPARC.

\vspace{-0.3cm}
\bibliographystyle{aa} 

\end{document}